%% file: paper.tex
\documentclass[conference]{IEEEtran}
\IEEEoverridecommandlockouts
\usepackage[english]{babel}
\usepackage[numbers]{natbib}
\usepackage[acronym]{glossaries}
\usepackage{hyperref}
\usepackage[capitalise]{cleveref}
\usepackage{lipsum} 
\input{acro.tex}
\input{macros.sty}

\usepackage{cite}
\usepackage{amsmath,amssymb,amsfonts}
\usepackage{algorithmic}
\usepackage{graphicx}
\usepackage{textcomp}
\usepackage{xcolor}
\def\BibTeX{{\rm B\kern-.05em{\sc i\kern-.025em b}\kern-.08em
    T\kern-.1667em\lower.7ex\hbox{E}\kern-.125emX}}
\begin{document}

\title{
Charting 5G Energy Efficiency: Flexible Energy Modeling for Sustainable Networks
\\
}


\author{\IEEEauthorblockN{Anderson L. de Araujo\IEEEauthorrefmark{1},
Luc Deneire\IEEEauthorrefmark{1},
Guillaume Urvoy-Keller \IEEEauthorrefmark{1}, 
André L. F. de Almeida\IEEEauthorrefmark{2}}
\IEEEauthorblockA{\IEEEauthorrefmark{1}Université Côte d'Azur, I3S, Sophia Antipolis, France}
\IEEEauthorblockA{\IEEEauthorrefmark{2}Department of Teleinformatics Engineering, UFC, Fortaleza, Brazil}%
}

\maketitle

\begin{abstract}
Despite the rapid advancements in 5G technology, accurately assessing the energy consumption of its Radio Access Networks (RANs) remains a challenge due to the diverse range of applicable technologies and implementation solutions. To estimate the energy consumption in 5G networks, this study  proposes a new method to model the energy of the baseband of the RANs, along with an end-to-end model inspired by the literature. The objective is to design a versatile energy model capable of estimating RAN-specific energy consumption, encompassing both mobile terminals and the physical layer (PHY) of base stations. The core of the model lies in the estimation of the number of operations done by PHY-layer algorithms.  This method is compared with the estimation of the number of cycles (and energy per cycle) used by a specific implementation (here a Matlab code ported on an Intel target). This enables to assess the model with the estimation of energy consumed on a real target. Overall, the results highlight the need for improved modeling techniques to better match the energy consumption patterns observed in the simulations. The main contribution of this study is a first step towards a flexible energy model with smaller granularity that can be used to compare the energy used  for multiple applications and in different contexts, providing a comprehensive tool for assessing and optimizing 5G network energy consumption.
\end{abstract}


\section{Introduction}


Over the past decade, the mobile communication sector has witnessed remarkable growth, fueled by the increasing demand for data traffic and services \citep{ericson2022mobrep}. This surge in activity has led to a corresponding rise in energy consumption. Mobile networks typically consist of the Radio Access Network (RAN), which includes user equipment (UE) and base stations (BS), along with the network core. Notably, BSs can account for up to $80\%$ of the energy consumption in mobile networks' operational expenditure (OPEX) \citep{nokia-wp}. Consequently, academia and industry have made concerted efforts to develop more environmentally friendly mobile networks.


Research efforts have focused on developing power models for base stations (BS) due to their significant energy consumption (\citep{auer2011much} and \citep{desset2012flexible} present models based on 4G networks). In "How much energy is needed to run a wireless network?" \citep{auer2011much}, the BS is broken down into its constituent components: \gls{BB}, \gls{RF}, \gls{PA}, and overhead (OH, related to cooling and power supply). Desset et al. \citep{desset2012flexible} further refine this model by dissecting the BBU and RF components into subcomponents and incorporating their operations as model inputs. Yan et al.\citep{yan2019modeling} introduce a model to estimate energy consumption for mobile services based on different application types. By characterizing mobile services according to network topology segments, the authors estimate energy consumption for each part of the topology (\gls{UE}, \gls{BS}, wireline core, and \gls{DC}) to determine the total energy consumed during mobile service usage.


The development of the latest mobile generation, driven by the need for higher throughput and lower latency, prioritized ultra-lean design principles, to minimize non-data-related transmissions. This led to the creation of 5G \gls{NR}, aimed at improving energy efficiency compared to 4G \gls{LTE}. The survey by  \citet{lopez2022survey} explores the current state of research on 5G energy efficiency. They review various power consumption models for distributed and centralized \gls{RAN} architectures and discuss key energy efficiency technologies, highlighting their benefits and challenges through detailed examinations of implementations and operational aspects.

\citet{williams2022energy} point out gaps in studies related to 5G energy consumption, emphasizing the need for openly accessible, vetted, and transparent assessments of the entire network. They underscore the importance of considering not only direct energy savings but also indirect and rebound effects when evaluating 5G's energy-saving potential.

 \citet{larsen2023toward} gather research efforts aimed at minimizing the energy consumption of future mobile networks. They explore various approaches, such as RAN architecture enhancements, technological improvements, and network sharing among operators. The authors discuss the feasibility of these approaches in real-world scenarios, highlight their current state of development, and propose guidelines to promote the environmental sustainability of NR and subsequent mobile generations.


Regarding 5G (\gls{NR}), several studies have investigated its energy and power consumption. \citet{xu2020understanding} and \citet{narayanan2021variegated} examined the power consumption of \glspl{UE}, smartphones, when using 5G networks compared to 4G \gls{LTE}. \citet{tombaz2015energy} evaluated the power consumption of a 5G system incorporating massive beamforming and ultra-lean design principles, employing a power model based on resource utilization. \citet{yu2015joint} addressed an energy-efficient resource allocation problem for joint downlink and uplink transmission with \gls{ca} through optimization techniques. \citet{fu2020end} adapted a power model originally described by \citet{desset2012flexible} for 5G \glspl{BS}, \glspl{gNB}, along with the power coonsumption of \gls{Li-Fi} technologies and \gls{mmWave} \glspl{IAP} and their respective power consumption levels.

Additionally, given the diverse range of technologies applicable to 5G and the various implementation solutions available, conducting a robust energy assessment of the Radio Access Network (RAN) through a power model requires extensive data collection and experimental studies to accurately identify the relevant parameters specific to real-world 5G equipment. Thus, the question is : is it possible to design a versatile power model to capture the range of technologies applied to 5G to estimate the \gls{RAN} specific power consumption encompassing both the mobile terminal and the PHY part of the base station? 

The objective of this study is to contribute to the development (and to enhance) of an end-to-end energy/power model to estimate the amount of power consumed when using 5G networks. The model is based on the algorithms and protocols used by the 5G \gls{NR} \gls{PHY} and their costs per operation. The core of the model involves estimating the number of operations performed by \gls{PHY} protocols in both base stations and user equipment. Overall, the results highlight the need for improved modeling techniques to better match the energy consumption patterns observed in the simulations, such as the adoption of optimized matrix multiplication algorithms. The model itself is compared to energy estimation on a real target (here an Intel target with code generated by Matlab).  The main contribution of this study is a first step to a flexible energy model with smaller granularity that can be extended for multiple applications and different contexts, providing a comprehensive tool for assessing 5G network energy consumption.

The outline of the paper is as follows. \cref{SotA} introduces the related works done before. The model and simulation parameters are presented in \cref{Model}. The results are presented and discussed in \cref{Res}. Finally, \cref{Conc} concludes the paper.
 
\section{Related Works}\label{SotA}

In this section we highlight pertinent studies that have contributed to the state of the art and understanding of power consumption modeling of the \gls{RAN}. By synthesizing the current state of knowledge, we aim to highlight gaps, identify areas of consensus, and delineate the unique contributions of our work.

In the early 2010s, the \gls{EARTH} project (\citet{auer2011much}) analyzed the power consumption of \gls{BS} by breaking it down into components such as \gls{BB}, \gls{RF}, \gls{PA}, and OH. They found a linear relationship between the power consumption and the number of transceivers $N_{TRX}$, as well as the load level of the BS. This relationship was quantified in (\ref{eq:auer}), where $P_{in}$ represents the power consumed by the \gls{BS}, $P_{0}$ is the power consumed by a transceiver at minimum non-zero output power, $P_{sleep}$ denotes the power consumed in sleep mode by a transceiver, $P_{out}$ stands for the output power from the transceiver, and $\Delta_p$ represents the load-dependent slope.

 \begin{equation}\label{eq:auer}
 P_{in} = \begin{cases}
            N_{TRX} \left( P_{0} + \Delta_p P_{out} \right), &\text{if $0 < P_{out} < P_{max}$}\\
          
            N_{TRX}P_{sleep},  &\text{if $P_{out} = 0$}
\end{cases}
\end{equation}

In \citep{desset2012flexible}, the \gls{EARTH} project went deeper on the power breakdown of the \gls{BS} components stated in (\ref{eq:desset}), specially regarding the \gls{BB} and \gls{RF}, where the authors defined a dependency on the components and operations within these unities.

\begin{equation}\label{eq:desset}
    P_{BS} = P_{BBU} + P_{RF} + P_{PA}  + P_{OH}
\end{equation}

\citet{yan2019modeling} assessed the \gls{RAN} energy consumption by modeling the \gls{LTE} energy consumption by the sum of the segments of the network, such as the \gls{UE}, \gls{BS}, wireline core, that connects the \gls{RAN} to datacenters, and \gls{DC}, as in (\ref{eq:yan}).

\begin{equation}
    \label{eq:yan}
    E_{total} = E_{UE} + E_{BS} + E_{wireline} + E_{DC}
\end{equation}

Regarding the \gls{BS} energy consumption, denoted by $E_{BS}$, it is established a dependency on the number of resource elements used for data transmissions and control signaling.

In \citep{yu2015joint}, the authors developed a model for the power consumption of a 5G system incorporating \gls{ca}. They found that the power consumed at the \gls{BS} consists of both static and dynamic components, both of which are influenced by the system's load level. The model, described in (\ref{eq:yu}), considers parameters such as $P_{TX_{j}}$, $P^{CA}_{CP_{j}}$, and $P^{CAi}_{CP}$, representing the effective transmit power consumed by the $j$-th \gls{CC}, the variable and static power consumed by shared hardware components across \glspl{CC}, respectively. Additionally, $N{CC}$ and $B_{j}$ denote the number of \glspl{CC} and the bandwidth of the $j$-th \gls{CC}.

\begin{equation}
    \label{eq:yu}
    P_{BS} = \sum^{N_{CC}}_{j = 1}  (P_{TX_{j}} + B_{j} P^{CA}_{CP_{j}}) + P^{CAi}_{CP}
\end{equation}

Regarding \citep{tombaz2015energy}, the authors design a power model for 5G systems while incorporating \gls{mMIMO} on the link budget. The power consumed at the \gls{BS}, described in (\ref{eq:tombaz}), is proportional to the number of sectors $N_{s}$ in the \gls{BS}, and encompasses three scenarios: i) during transmission, including the transmit power per sector $P_{tx}^{s}$ adjusted by the \gls{PA} energy efficiency $\eta_{PA}$, the number of \gls{RF} chains $N$ and the additional digital and \gls{RF} processing per antenna $P_{C}$, along the baseline power consumption of each sector $P_{B}$; ii) in the absence of transmission, and no cell \gls{DTX}, then only the baseline power consumption is accounted; iii) in the absence of transmission, but cell \gls{DTX} is implemented, the the power consumed by the sector is a fraction of the baseline power, defined by the cell \gls{DTX} factor $\delta$.

\begin{equation}
    \label{eq:tombaz}
    P_{BS} = N_s \times \begin{cases}
                        \cfrac{P_{tx}^{s}}{\eta_{PA}} + N P_{C} + P_{B}, &\text{if $P_{tx}^{s} >0$}\\
                        P_{B} , &\text{if $P_{tx}^{s} =0$, no \gls{DTX}}\\
                        \delta P_{B} , &\text{if $P_{tx}^{s} =0$, and \gls{DTX}}
    \end{cases}
\end{equation}

\citet{fu2020end} tailored a power model based on \ref{eq:desset} where the \gls{BS} was composed of multiple nodes. For each node, the \gls{BB} and the \gls{RF} power consumption is a function of the operation complexity, $\mathcal{Q}$ measured by \gls{GFOS}, for operations within these components, adjusted by the technology-dependent factor $\rho$, measured in Giga floating-point operations per Watt (GOP/W). The \gls{BB} power consumption is described generically in \ref{eq:fu_bb}, where is presented the encoding, network and control operations, and the power consumed is a function of the number of beamforming $L$. The \gls{RF} power consumption is described in a similar way, in (\ref{eq:fu_rf}), but with the operations within the \gls{RF}, such as modulation, mixer, \gls{vga}, \gls{lna}, ADC and clock , and proportional to the number of antenna elements $M$.

\begin{equation}
    \label{eq:fu_bb}
    P_{BB} = L(\mathcal{Q}_{enc} + \mathcal{Q}_{net} + \mathcal{Q}_{ctrl})/\rho
\end{equation}

\begin{equation}
    \label{eq:fu_rf}
     P_{RF} = M(\mathcal{Q}_{mod} + \mathcal{Q}_{mix} + \mathcal{Q}_{vga} + \mathcal{Q}_{lna} + \mathcal{Q}_{adc})/\rho + \sqrt{M} \mathcal{Q}_{clk} / \rho
\end{equation}

\section{Method}\label{Model}

In this section, we develop the  energy model, inspired by \citep{fu2020end}. It is based on the algorithms and protocols used by the \gls{NR} \gls{PHY}, depicted on the \gls{3GPP} technical specifications, and their costs per operation. The core is the estimation of the number of operations and the number of CPU cycles at the \gls{BB}, generated from the protocols from the \gls{PHY} downlink \gls{BS} and the \gls{UE}  and, from that, calculate the energy consumed for transmission for a given scenario. 


 The model is composed by eight blocks, each one is a group of algorithms done for the processing of the signal. On the \gls{BS} side, there are: \gls{crc} attachment, \gls{CB} segmentation and \gls{crc} attachment, channel coding, rate matching, in block A; \gls{CB} concatenation, scrambling, modulation, layer mapping, in block B; antenna port mapping, mapping to to \gls{vrb}, mapping from \gls{vrb} to \gls{PRB}, in block C; \gls{OFDM}, and \gls{cp} addition, in block D. On the \gls{UE} side, there are \gls{cp} remove, \gls{OFDM} demodulation, in block E; channel estimation, in block F; layer demapping, descrambling, in block G; and rate recovery, channel decoding, desegmentation, and \gls{crc} decoding in block H. The notation $\mathcal{O}$ represents the number of operations estimated by the model. It is assumed that the addition and operations for memory vectors have the same cost in terms of \gls{micro-op} for the matrix inversion, for the \gls{ls} estimation, the \gls{svd} and the \gls{fft} used to compute the \gls{dft} algorithms, . 
 

 The validation of this model is done by comparison with a MATLAB simulation, where the \gls{NR} \gls{PHY} is implemented using the 5G toolbox. The \gls{soc} Blockset toolbox in MATLAB is used to get the number of operations at each one of the Blocks described in this section, from a \texttt{xlsx} file generated by it. The main inputs of this simulation are the number of slots, the desired \gls{snr}, the modulation, the code rate, the subcarrier spacing, the gridsize, and the number of antennas at the transmitter and at the receiver. 

Once the operations are quantified and sorted into the different types, the next step is to map them into assembly instructions, to go into a lower level on the CPU processor and quantify the number of cycles necessary to perform those operations. \citet{fog1996optimization} presents a mapping from the assembly instructions and the number of cycles needed to perform the required operation in a specific type of memory operand, used for both model and simulation. Then, once the total number of cycles is acquired, the total energy consumed is proportional to the number of cycles. The energy consumed per cycle $\epsilon$, as used in \citet{hao2018energy} and described in (\ref{eq:energy/cycle}), depends on the squared clock frequency of the processor $f_{pc}$, and an energy coefficient $\kappa$ .

\begin{equation}
    \label{eq:energy/cycle}
    \epsilon = \kappa f_{pc}
\end{equation}

\subsection{Block A}

The most relevant algorithms, in terms of the number of operations, are the \gls{crc} attachment, \gls{CB} segmentation and \gls{crc} attachment, and channel coding. The base graph selection, rate matching, and \gls{CB} concatenation do not contribute significantly compared to these algorithms. The operations for \gls{crc} attachment depend on the input size. Let $A$ be the number of input bits. In each iteration of the \gls{crc} algorithm using the slice-by-4 method, $p$ bits are read per step. The number of AND, XOR, and shift operations are performed equally, as described in (\ref{eq:crc}). For \gls{CB} segmentation and \gls{crc} attachment, the segmentation requires nine operations per \gls{TB}, and the \gls{crc} attachment is done for each \gls{CB}. The total number of operations, for all the \glspl{crc} calculation for the \glspl{CB}, is approximated to the sum of all \gls{CB} sizes. If the \gls{TB} size is $A$, the sum of all \gls{CB} sizes $B$ is defined as $B = A + L C$, where $L$ is the \gls{crc} prefix and $C$ is the number of \glspl{CB}.


\begin{equation}
    \label{eq:crc}
    \mathcal{O}_{crc} = 5\left \lfloor\frac{A}{p} \right\rfloor + 1
\end{equation}

Regarding channel coding for the data channel, the \gls{ldpc} algorithm is applied for each \gls{CB}. Let $K$, $Z_{i,j}$, $n_1$, $N_{BG}$, $M_{BG}$, and $N_{cCB}$ be the maximum \gls{CB} size, the \gls{ldpc} lifting size, the number of non-null elements in a base-graph matrix, the number of rows, and the number of columns in a base-graph matrix, and the size of the coded \gls{CB}, defined by the technical specifications, respectively.

The process, for each \gls{CB}, is depicted into validation of non-null input bits, described in (\ref{eq:ldpc-val}), replacements on the base graph matrix, described in (\ref{eq:ldpc-rep}), the calculation of the modulo operator, which has the cost of a division operation, described in (\ref{eq:ldpc-n1}) a large matrix product, described in (\ref{eq:ldpc-mp}), and setting of values into the output \gls{CB}, described in (\ref{eq:ldpc-set}). The total number of instructions generated by the channel coding is equal to the sum of the instructions from (\ref{eq:ldpc-val}) to (\ref{eq:ldpc-set}) times the number of \glspl{CB} $C$.

\setlength{\arraycolsep}{0.0em}
\begin{eqnarray}
\mathcal{O}_{LDPC-val} = && 2(K - 2Z_{i,j}) \label{eq:ldpc-val}\\
\mathcal{O}_{LDPC-rep} = && N_{BG} \cdot M_{BG}\label{eq:ldpc-rep}\\
\mathcal{O}_{LDPC-n1} = && n_1 \label{eq:ldpc-n1}\\
\mathcal{O}_{LDPC-mp} = && N_{BG}\cdot Z_{i,j}(2 M_{BG} \cdot Z_{i,j} -1)\label{eq:ldpc-mp}\\
\mathcal{O}_{LDPC-set} = && N_{cCB} + 2 Z_{i,j} - K\label{eq:ldpc-set}
\end{eqnarray}
\setlength{\arraycolsep}{5pt}

\subsection{Block B}

In this block, three procedures are applied to each \gls{TB} (also named codeword): scrambling, modulation and layer mapping. Scrambling: This involves an XOR operation between the codeword of size $M_{cw}$ and a scrambling sequence of the same size. The scrambling sequence is generated using 5 XOR operations, resulting in a total of 6$M_{cw}$ XOR operations for scrambling, in (\ref{eq:scram}). Modulation: Performed at the \gls{RF} level, andso, not entirely taken in account by this model, this is considered a generic operation, such as a table look-up for each modulation symbol, in (\ref{eq:mod+lm}). Layer Mapping: This involves shifting the modulated symbols into the layers available for transmission, requiring $N_{symbols}$ shift operations.

\setlength{\arraycolsep}{0.0em}
\begin{eqnarray}
\mathcal{O}_{scmb} = && 6 M_{cw} \label{eq:scram}\\
\mathcal{O}_{mod} = && N_{symbols}\label{eq:mod+lm}
\end{eqnarray}
\setlength{\arraycolsep}{5pt}

\subsection{Block C}

In this block, the main step is the \gls{apm}, where the precoding is done based on the \gls{csi} report, as well as the mapping from the $v$ layers to the $P$ antenna ports indicateb by the \gls{csi} report. The steps mapping to \gls{vrb} and mapping from \gls{vrb} to \gls{prb} can be omitted since the operations done here are small compared to the \gls{apm}. The \gls{apm} can be resumed in the calculation of the precoding matrix by the application of the \gls{svd} on the information present on the \gls{csi} report, and the matrix multiplication of the precoding matrix and the block of vectors $[x^{(0)}(i), \cdots , x^{(v-1)}(i)]^T, i = 0, 1, \cdots M_{symb}^{layer} - 1$, where $M_{symb}^{layer} = N_{symbols}/v$. The precoding matrix has dimensions $P \times v$, and each Block of vectors has dimensions $v \times 1$. 

The \gls{svd} of a matrix with dimensions $m \times n$, has a around  $2 mn^2 + n^3 + n + mn$ \glspl{flop}. Then, this block, has a total of $2 Pv^2 + v^3 + v + Pv$ \glspl{flop} from the \gls{svd} calculation, and $2Pv - P$ \glspl{flop} from the matrix calculation, per $M_{symb}^{layer}$, resumed in (\ref{eq:apm}).

\begin{equation}
    \label{eq:apm}
    \mathcal{O}_{apm} = M_{symb}^{layer}(2 Pv^2 + v^3 + v + Pv + 2Pv - P)
\end{equation}

\subsection{Blocks D and E}

The Blocks D and E are described together as Block D represents the \gls{OFDM} modulation and \gls{cp} addition, while Block E represents the reverse operations, \gls{OFDM} demodulation and \gls{cp} remove. As the \gls{cp} related operations are not significant compared to the \gls{OFDM} (de)modulation, they are not taken in account.

 The modulation is done by the \gls{ifft} application at the transmitter side on the resource grid of dimensions $N_f \times g \times N_{Tx}$, where $f, g$, and $N_{Tx}$ are the number of subcarriers, the number of \gls{OFDM} symbols and the number of transmit antennas. While the demodulation is done by the \gls{fft} application on the received signal at the receiver side to recover the same grid. The number of operations done in this step is defined by the application of the Radix-2 (i)FFT algorithm, with \gls{fft} size equal to $N_{FFT}$. It leads to the number of operations in (\ref{eq:fft}), where $N_{FFT} > N_f$. As Block E is the inverse operation, \gls{fft}, the number of operations are the same.
 
 \begin{equation}
    \label{eq:fft}
    \mathcal{O}_{FFT} = 5 g N_{Tx} N_{FFT} \log_2 N_{FFT}
\end{equation}



\subsection{Block F}
This block is responsible for the channel estimation on the \gls{UE} side, where the channel is estimated by using the pilot symbols and \gls{ls} estimation, and \gls{mmse} equalization is applied to data symbols.

Inspired by \citep{barhumi2003optimal}, the channel estimation by the \gls{ls} method, per reference signal port (the same as layers for Block C) and per receiver antenna is obtained by the product of the pseudo-inverse of the pilot training symbols matrix and the received signal at the receive antenna. Let $L$, $g$, $f$ and $f_p$ be the maximum channel length, the number of \gls{OFDM} symbols in a slot, the total number of \gls{OFDM} subcarriers and the \gls{OFDM} subcarriers with pilots. To obtain the pseudo-inverse of $\tilde{\mathsf{A}} \in \mathbb{C}^{g f_p \times L N_t}$, the matrix composed by the pilot sequences, it is necessary to perform a matrix product, followed by a matrix inversion, and a last matrix product. The first matrix product has $(L N_t)^2(2gK_p-1)$, the inversion operation, using the LU decomposition, has $(L N_t)^3$ operations, and the final matrix product has $(g K_p L N_t)(2LN_t-1)$ operations. The estimation of the $N_r$ channels perceived by the $v$ reference signals, each one experienced  by a receiver antenna, leads to (\ref{eq:ls}).

\begin{equation}
\begin{split}
    \label{eq:ls}
    \mathcal{O}_{LS} = v N_r \left[(L N_t)^2(2gK_p-1) + (L N_t)^3\right. \\ + \left.(g K_p L N_t)(2LN_t-1)\right]
    \end{split}
\end{equation}
Regarding the equalization, the \gls{mmse} algorithm, its operations are dominated by the \gls{svd} calculation of the channel matrix, and the application of that equalization, for each subcarrier. Consider the system described in (\ref{eq:sys}), where $Y(f) $ and $ Z(f) \in \mathbb{C}^{Nr \times g}, H(f) \in \mathbb{C}^{N_r \times N_t}$, and $ X(f) \in \mathbb{C}^{N_t \times g}$ are, respectively the received signal and the noise at subcarrier $f$, the channel matrix and the transmitted symbols at subcarrier $f$. The svd of $H(f)$ leads to $2(N_r)(N_t)^2 +(N_t)^3 + N_r + N_r N_t$ \glspl{flop} per subcarrier. 

\begin{equation}
    \label{eq:sys}
     Y(f) = H(f) X(f) + \eta(f)
\end{equation}

Given the \gls{svd} of $H(f) = U(f) \Sigma(f) V^{H}(f)$, where $ U(f) \in \mathbb{C}^{N_r \times N_r}, \Sigma(f) \in \mathbb{R}^{N_r \times N_t}$ and $V(f) \in \mathbb{C}^{N_r \times N_t}$, the equalized symbols is described in (\ref{eq:mmse}), where $\Xi = [ \Psi_{N_t \times N_t} \quad 0_{(N_r-N_t) \times N_t}]^{T}$ and $\Psi$ is described in (\ref{eq:psi}), where $\sigma^{2}_{n}$ is the noise variance, and $\sigma_i, i = 1, \cdots, N_t$ is the $i$-th singular value in $\Sigma(F)$.
\begin{equation}
    \label{eq:mmse}
     \Hat{X}(f) = V(f) \Xi(f) U^{H}(f) Y(f)
\end{equation}

\begin{equation}
    \label{eq:psi}
    \Psi = \mathrm{diag}\left( \cfrac{\sigma_{i}(f)}{\sigma_{i}^{2}(f) + \sigma_{n}^{2}(f)} \right) 
\end{equation}
 For each subcarrier, the construction of $\Psi$ requires 3 operations to be done per singular value, and there are 3 matrices multiplications with $N_t N_r (2N_t - 1)$, $N_t N_r (2N_r - 1)$, and $N_t g (2N_r - 1)$ operations, respectively, leading to (\ref{eq:eq}) total operations.
 
\begin{equation}
 \begin{split}   \label{eq:eq}
    \mathcal{O}_{eq} = 2 N_r N_t^2 + N_r^3 + N_r + N_r N_t + N_f \left[ 3N_t 
     \right. \\ \left.+ N_t N_r (2N_t - 1) + N_t N_r (2N_r - 1)  + N_t g (2N_r - 1) \right]
    \end{split}
\end{equation}

\subsection{Block G}
 This block is the reverse block of block B, and so the operations here ate the inverse of the ones done at block B: layer demapping, symbol demodulation, and symbol descrambling. The number of operations done in this block is described by (\ref{eq:scram}) and (\ref{eq:mod+lm}), as well.

\subsection{Block H}
 The last block is composed by the rate recovery, channel decoding, desegmentation, and \gls{crc} decoding, where the most relevant operations come from the \gls{ldpc} decoding, desegmentation, where a \gls{crc} decoding is done over the \glspl{CB} and a final \gls{crc} decoding done over the \gls{TB}.

 Regarding the \gls{ldpc} decoding, it is done by a message passing algorithm and using \glspl{llr}, and the Min-Sum algorithm. Let $N$, $W$, $|\mathcal{N}(w)|$, and $|\mathcal{W}(n)|$ be, respectively, the number of variable nodes (i.e.,the number of bits on the coded \gls{CB}), the number of check nodes (i.e.,the number of redundancy bits added at the encoding), the number of variable nodes connected to the check node $w$, and the number of check nodes connected to  variable node $n$. The decoding process is done in four steps: initialization, done at the beginning of decoding with $N$ quotients and $N$ log calculations; horizontal step, done once per iteration, with $w|\mathcal{N}(w)|$ products; vertical step, done once per iteration, with $N|\mathcal{W}(n)|$ additions; and decision step, done once per iteration, with $N(|\mathcal{W}(n)|+1)$ additions and $\sum_{w = 1}^{W}|\mathcal{N}(w)|$ XOR products.

Similarly to the \gls{BS} side, the \gls{crc} decoding is done for the \gls{TB} and for the \gls{CB}. It is done by a comparison between the received \gls{crc} header and a new header computed at the receiver, using the received \gls{TB}. If they are equal, then, the data is not corrupted, otherwise, there is a failure during transmission. Hence, the number of operations (AND, XOR and shifts) done at the decoding for \gls{TB} and \glspl{CB} is equal to the number of operations done at the encoding, plus the comparison.



\section{Results and Discussion}\label{Res}
In this section we present the findings of the study to develop a 5G \gls{NR} end-to-end energy model based on the \gls{PHY}-layer protocols and algorithms. The model results are under the tag Model, while the validation results are under the tag Measurements.

During the simulation, functions other than the ones at the 5G and Communications System toolboxes are used, such as auxiliary functions for code generation. The file obtained from the simulation is composed by three reports per function analyzed. One is the Path aggregated report, used to identify the desired toolboxes for measurements, other is the Operator aggregated report, not used in this validation, and the last one is the Operator detailed report, where the information necessary for validation is obtained.

The reports are processed using the \texttt{pandas} library from python. First, the MATLAB auxiliary functions are excluded to avoid noise in the energy estimation, by filtering the data present on the Path aggregated report. Then, the Operator detailed report is used to get the necessary information for the estimation, including data type, operation, and count of operations. There are mainly seven types of data: logical, int32, and double (which means a double precision floating point), in both scalar and vector flavours, and struct as well. The operators are translated into assembly instructions and then into \glspl{micro-op} using the data from \citep{fog1996optimization} (data updated in 2022), based on the type of memory used. It is assumed that scalars are stored in register operands and vectors are stored in memory operands (logical and integer vector are assumed to be stored in mmx registers, and double vectors to be stored in xmm registers). 

The parameters from each block are collected from the simulation to be used as inputs for the estimation model . The model returns the number of operations, and the operations if not defined in \cref{Model} are assumed to be a \gls{flop}, i.e., an addition and a multiplication. The datatypes of Blocks A, G, scrambling and modulation inputs, from Block B, and descrambling input, from Block G, are assumed to be integer or logical while the other algorithms ad blocks are assumed to have a double type inputs. Regarding the Block H, the number of variable nodes connected to a check node, the number of check nodes connected to a variable node and the number of iterations is assumed to be a constant, for sake of simplicity for this initial study.

\begin{figure}[htbp]
\centerline{\includegraphics[scale=0.6]{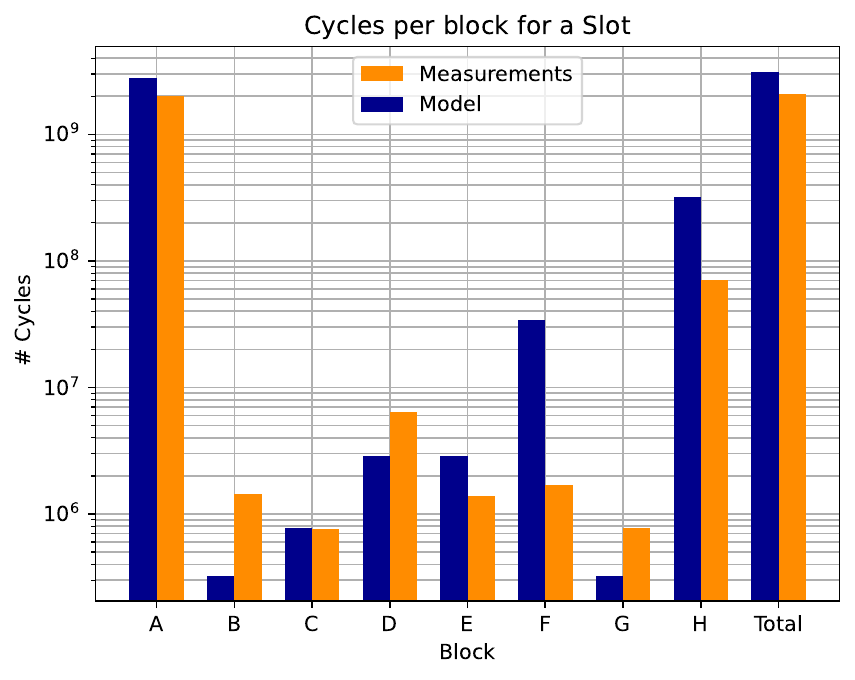}}
\caption{Number of cycles per block and the total number for a single slot transmission on downlink.}
\label{fig}
\end{figure}

Fig.~\ref{fig} is the graphic representation of the results obtained from the Measurements and Model estimation for one slot with a single codeword, 10 db of \gls{snr}, 15 KHz subcarrier spacing, 16 \gls{qam}, coding rate equal to $490/1024$, four antennas on both transmitter and receiver, and two layers. The value considered for $\kappa$ is $10^{-25} J\cdot s^2$, the same value used in \citet{hao2018energy}, and $f_p$ equal to 2.1 GHz. The vertical axis is in logarithmic scale to allow the visualization of all the blocks. Each block has two bars, one for the Model, in blue, and other for the Measurements, in orange. In the blocks A, E, F, and H, the model presents an overestimation of the amount of energy consumed, compared to the simulations, while in the blocks B, C, D, and G, the model presents an underestimation of the energy consumed. 

The results for the energy estimation by the model, compared to the simulations results obtained from MATLAB, are similar (discrepancies are explainable). The blocks with large matrix multiplication, like in blocks A and F, can be overestimated due to the model used for matrix multiplication, a schoolbook based approach, that can be too greedy, specially in the cases of sparse matrices, like the base graph matrix for channel coding, while MATLAB uses optimized algorithms. The same happens for the matrix inversion. In other blocks like B and G, the underestimation can be explained by the fact that while the model assume that most of its energy consumption comes from the \gls{RF} chain, and so, does not consider it, while the measurements, in MATLAB, does compute all the modulation operations, and so, the validations results contains both RF and BBU parts.

\begin{figure}[htbp]
\centerline{\includegraphics[scale=0.6]{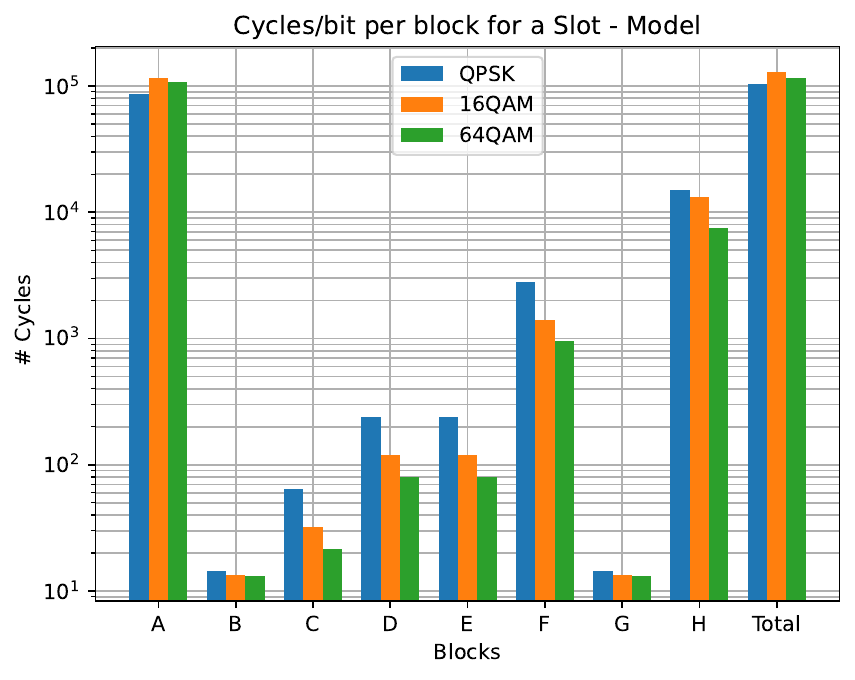}}
\caption{Number of cycles per bit estimated by the model for the blocks and total number of cycles of modulations QPSK, 16QAM, and 64QAM for a single slot transmission on downlink.}
\label{fig-model}
\end{figure}

\begin{figure}[htbp]
\centerline{\includegraphics[scale=0.6]{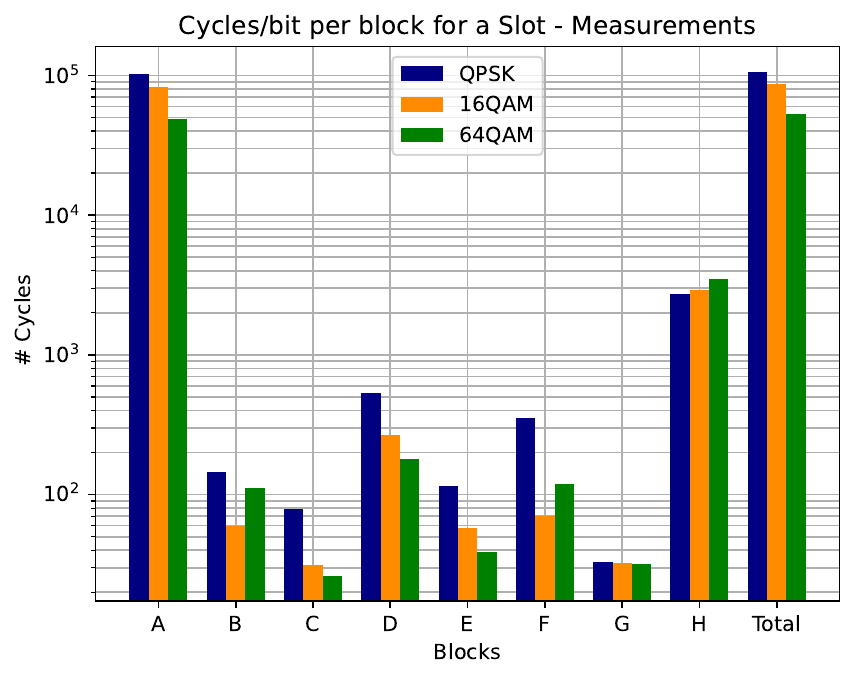}}
\caption{Number of cycles per bit estimated by the measurements for the blocks and total number of cycles of modulations QPSK, 16QAM, and 64QAM for a single slot transmission on downlink.}
\label{fig-sim}
\end{figure}

Figs.~\ref{fig-model} and \ref{fig-sim} are the results of results obtained in terms of cycles per bit by changing only the modulation to QPSK, 16 QAM and 64 QAM for model and measurements, respectively. Once the total number of cycles per block is acquired, it is divided by the number of bits sent over the transmission. In both Figs, is possible to see that the higher the modulation, the less the number of cycles required to process and transmit them, in accord which is expected. However, in the model, for Block A, the same behavior is not seen, which requires further investigations. 

As the available results are in a micro-scale, i.e., at a slot level, a direct comparison with other methods cannot be established up to now. The results indicate that the model is globally in agreement with the simulations, but we need to take further optimized algorithms (e.g optimized matrix operation algorithms) into account. 
The model presents a first step towards a method for smaller granularity that can be extended for multiple applications and different contexts, hence opening the possibility to compare the real impact of different algorithmic and architectural choices in the implementation of 5G.









\section{Conclusion}\label{Conc}

By proposing a method based on the number and type of computational operations performed at the 5G \gls{PHY} for a downlink transmission and reception, we address the  development of a versatile energy model applied to 5G to estimate the RAN specific energy  consumption encompassing both the mobile terminal and the \gls{PHY} part of the base station. We design a first version of the model based on eight blocks, whose implementation based complexity estimations are similar to the operations based model. Further enhancements on the model are necessary for the application of this model in a larger scale and different contexts.

\bibliographystyle{IEEEtranN}
\bibliography{refs}  

\end{document}

%% file: acro.tex
\newacronym{ICT}{ICT}{Information and Communication Technology}
\newacronym{LTE}{LTE}{Long Term Evolution}
\newacronym{3GPP}{3GPP}{Third Generation Partnership Project}
\newacronym{RAN}{RAN}{Radio Access Network}
\newacronym{CN}{CN}{Core Network}
\newacronym{NR}{NR}{New Radio}
\newacronym{SDO}{SDO}{Standards Developing Organization}
\newacronym{eMBB}{eMBB}{enhanced Mobile Broadband}
\newacronym{uRLLC}{uRLLC}{ultra-Reliable Low-Latency Communication}
\newacronym{mMTC}{mMTC}{massive Machine Type Communication}
\newacronym{IoT}{IoT}{Internet of Things}
\newacronym{VR}{VR}{virtual Reality}
\newacronym{MIMO}{MIMO}{Massive Input Massive Output}
\newacronym{QoS}{QoS}{Quality of Service}
\newacronym{QoE}{QoE}{Quality of Experience}
\newacronym{BS}{BS}{base station}
\newacronym{UE}{UE}{user equipment}
\newacronym{LBT}{LBT}{listen-before-talk}
\newacronym{OFDM}{OFDM}{Orthogonal Frequency-Division Multiplexing}
\newacronym{mmWave}{mmWave}{milimeter wave}
\newacronym{NSA}{NSA}{non-stand-alone}
\newacronym{SA}{SA}{stand-alone}
\newacronym{EPC}{EPC}{Evolved Packet Core}
\newacronym{CUPS}{CUPS}{Control and User Plane Separation}
\newacronym{NFV}{NFV}{Network Function Virtualization}
\newacronym{UCA}{UCA}{Universté Côté d'Azur}
\newacronym{ubinet}{ubinet}{ubiquitous network}
\newacronym{I3S}{I3S}{\textit{ Laboratoire d'Informatique, Signaux et Systèmes de Sophia Antipolis}}
\newacronym{RF}{RF}{radio frequency}
\newacronym{PA}{PA}{power amplifier}
\newacronym{BB}{BBU}{baseband unity}
\newacronym{GSM}{GSM}{Global System for Mobile communication}
\newacronym{UMTS}{UMTS}{Universal Mobile Telecommunications System}
\newacronym{RRC}{RRC}{Radio Resource Control}
\newacronym{B5G}{B5G}{beyond 5G}
\newacronym{IAP}{IAP}{indoor access point}
\newacronym{Li-Fi}{Li-Fi}{Light-Fidelity}
\newacronym{BMAA}{BMAA}{building mounted antenna array}
\newacronym{MBS}{MBS}{macro base station}
\newacronym{MBSALA}{MBSALA}{antenna arrays}
\newacronym{MRC}{MRC}{Maximum Ratio Combining}
\newacronym{LoS}{LoS}{Line of sight}
\newacronym{PRB}{PRB}{Physical Resource Block}
\newacronym{RE}{RE}{ressouce element}
\newacronym{RMS}{RMS}{root mean square}
\newacronym{RSRP}{RSRP}{Reference Signal Received Power}
\newacronym{RTT}{RTT}{Round Trip Time}
\newacronym{UDP}{UDP}{User Datagram Protocol}
\newacronym{ACK}{ACK}{Acknolegment}
\newacronym{MAPE}{MAPE}{Mean Absolute Percentage Error}
\newacronym{DTR}{DTR}{Decision Tree Regression}
\newacronym{RRH}{RRH}{radio remote head}
\newacronym{RRU}{RRU}{radio remote unity}
\newacronym{GHG}{GHG}{green house gases}
\newacronym{AAUs}{AAUs}{active antenna unities}
\newacronym{MAC}{MAC}{media access control layer}
\newacronym{gNB}{gNB}{next generation Node B}
\newacronym{eNB}{eNB}{evolved Node B}
\newacronym{CRAN}{C-RAN}{Cloud Radio Access Network}
\newacronym{ORAN}{O-RAN}{Open Radio Access Network}
\newacronym{DC}{DC}{data center}
\newacronym{WL}{WL}{wireline core}
\newacronym{opex}{OPEX}{operation expenditure}
\newacronym{pdsch}{PDSCH}{physical downlink shared channel}
\newacronym{pdcch}{PDSCH}{physical downlink control channel}
\newacronym{pbch}{PBCH}{physical brodcast channel}
\newacronym{pusch}{PUSCH}{physical uplink shared channel}
\newacronym{pucch}{PUCCH}{physical uplink control channel}
\newacronym{prach}{PDSCH}{physical random access channel}
\newacronym{DAC}{DAC}{digital analog converter}
\newacronym{ADC}{ADC}{analog digital converter}
\newacronym{VGA}{VGA}{variable gain atenuattor}
\newacronym{OH}{OH}{overhead}
\newacronym{OAI}{OAI}{OpenAirInterface}
\newacronym{DU}{DU}{distributed unity}
\newacronym{CU}{CU}{central unity}
\newacronym{PHY}{PHY-layer}{physical layer}
\newacronym{GFOPS}{GFO/S}{Giga floating-point operations per second}
\newacronym{crc}{CRC}{cyclic redundancy check}
\newacronym{ldpc}{LDPC}{low-density parity check}
\newacronym{vrb}{VRB}{virtual resource block}
\newacronym{prb}{PRB}{physical resource block}
\newacronym{TTI}{TTI}{transmission time interval}
\newacronym{qam}{QAM}{quadrature amplitude modulation}
\newacronym{awgn}{AWGN}{additive white gaussin noise}
\newacronym{gop/w}{GFO/S}{Giga floating-point operations per watt}
\newacronym{TB}{TB}{transport block}
\newacronym{CB}{CB}{code block}
\newacronym{RB}{RB}{resource block}
\newacronym{IQ}{IQ}{in phase and quadrature}
\newacronym{qpsk}{QPSK}{quadrature phase shif keying}
\newacronym{csi}{CSI}{channel state information}
\newacronym{TS}{TS}{technical specification}
\newacronym{uci}{UCI}{uplink control information}
\newacronym{pi-bpsk}{$\pi/2$-BPSK}{$\pi/2$ binary phase shift keying}
\newacronym{tpmi}{TPMI}{transmit precoding matrix indicator}
\newacronym{papr}{PAPR}{Peak-to-Average Power Ratio}
\newacronym{fdma}{FDMA}{frequency division multiplexing access}
\newacronym{sc-fdma}{SC-FDMA}{single carrier frequency division multiplexing access}
\newacronym{dft}{DFT}{discrete Fourier transform}
\newacronym{ptrs}{PT-RS}{phase tracking reference signal}
\newacronym{fft}{FFT}{fast Fourier transform}
\newacronym{mmse}{MMSE}{Minimum Mean Squared Error}
\newacronym{ls}{LS}{Least Squares}
\newacronym{svd}{SVD}{Single Value Decomposition}
\newacronym{dmrs}{DM-RS}{demodulation reference signal}
\newacronym{zf}{ZF}{zero forcing}
\newacronym{llr}{LLR}{log-likelihood ratio} 
\newacronym{ifft}{iFFT}{inverse Fourier Fast Tramsform}
\newacronym{isi}{ISI}{Inter-Symbol Interference}
\newacronym{cp}{CP}{Cyclic Prefix}
\newacronym{ca}{CA}{Carrier Aggregation}
\newacronym{EARTH}{EARTH}{Energy Aware Radio and Network
Technologies}
\newacronym{CC}{CC}{component carrier}
\newacronym{mMIMO}{mMIMO}{massive Multiple Input Multiple Output}
\newacronym{DTX}{DTX}{discontinuous transmission}
\newacronym{GFOS}{GFO/S}{Giga floating-point operations per
second}
\newacronym{vga}{VGA}{variable gain amplifier}
\newacronym{lna}{LNA}{low-noise amplifier}
\newacronym{snr}{SNR}{signal-to-noise ratio}
\newacronym{soc}{SoC}{system on a chip}
\newacronym{micro-op}{$\mu$op}{micro-operation}
\newacronym{apm}{APM}{antenna port mapping}
\newacronym{flop}{FLOP}{floating-point operation}

%% file: paper.bbl
\begin{thebibliography}{16}
\providecommand{\natexlab}[1]{#1}
\providecommand{\url}[1]{#1}
\csname url@samestyle\endcsname
\providecommand{\newblock}{\relax}
\providecommand{\bibinfo}[2]{#2}
\providecommand{\BIBentrySTDinterwordspacing}{\spaceskip=0pt\relax}
\providecommand{\BIBentryALTinterwordstretchfactor}{4}
\providecommand{\BIBentryALTinterwordspacing}{\spaceskip=\fontdimen2\font plus
\BIBentryALTinterwordstretchfactor\fontdimen3\font minus
  \fontdimen4\font\relax}
\providecommand{\BIBforeignlanguage}[2]{{%
\expandafter\ifx\csname l@#1\endcsname\relax
\typeout{** WARNING: IEEEtranN.bst: No hyphenation pattern has been}%
\typeout{** loaded for the language `#1'. Using the pattern for}%
\typeout{** the default language instead.}%
\else
\language=\csname l@#1\endcsname
\fi
#2}}
\providecommand{\BIBdecl}{\relax}
\BIBdecl

\bibitem[Jonsson et~al.(2022)Jonsson, Lundvall, Möller, Carson, and
  Davies]{ericson2022mobrep}
P.~Jonsson, A.~Lundvall, R.~Möller, S.~Carson, and S.~Davies, ``Ericsson
  mobility report,'' Ericsson, Tech. Rep., nov 2022.

\bibitem[Nokia(2016)]{nokia-wp}
Nokia, \emph{5G network energy efficiency Massive capacity boost with flat
  energy consumption}, 2016.

\bibitem[Auer et~al.(2011)Auer, Giannini, Desset, Godor, Skillermark, Olsson,
  Imran, Sabella, Gonzalez, Blume, et~al.]{auer2011much}
G.~Auer, V.~Giannini, C.~Desset, I.~Godor, P.~Skillermark, M.~Olsson, M.~A.
  Imran, D.~Sabella, M.~J. Gonzalez, O.~Blume \emph{et~al.}, ``How much energy
  is needed to run a wireless network?'' \emph{IEEE wireless communications},
  vol.~18, no.~5, pp. 40--49, 2011.

\bibitem[Desset et~al.(2012)Desset, Debaillie, Giannini, Fehske, Auer,
  Holtkamp, Wajda, Sabella, Richter, Gonzalez, et~al.]{desset2012flexible}
C.~Desset, B.~Debaillie, V.~Giannini, A.~Fehske, G.~Auer, H.~Holtkamp,
  W.~Wajda, D.~Sabella, F.~Richter, M.~J. Gonzalez \emph{et~al.}, ``Flexible
  power modeling of lte base stations,'' in \emph{2012 IEEE wireless
  communications and networking conference (WCNC)}.\hskip 1em plus 0.5em minus
  0.4em\relax IEEE, 2012, pp. 2858--2862.

\bibitem[Yan et~al.(2019)Yan, Chan, Gygax, Yan, Campbell, Nirmalathas, and
  Leckie]{yan2019modeling}
M.~Yan, C.~A. Chan, A.~F. Gygax, J.~Yan, L.~Campbell, A.~Nirmalathas, and
  C.~Leckie, ``Modeling the total energy consumption of mobile network services
  and applications,'' \emph{Energies}, vol.~12, no.~1, p. 184, 2019.

\bibitem[L{\'o}pez-P{\'e}rez et~al.(2022)L{\'o}pez-P{\'e}rez, De~Domenico,
  Piovesan, Xinli, Bao, Qitao, and Debbah]{lopez2022survey}
D.~L{\'o}pez-P{\'e}rez, A.~De~Domenico, N.~Piovesan, G.~Xinli, H.~Bao,
  S.~Qitao, and M.~Debbah, ``A survey on 5g radio access network energy
  efficiency: Massive mimo, lean carrier design, sleep modes, and machine
  learning,'' \emph{IEEE Communications Surveys \& Tutorials}, vol.~24, no.~1,
  pp. 653--697, 2022.

\bibitem[Williams et~al.(2022)Williams, Sovacool, and
  Foxon]{williams2022energy}
L.~Williams, B.~K. Sovacool, and T.~J. Foxon, ``The energy use implications of
  5g: Reviewing whole network operational energy, embodied energy, and indirect
  effects,'' \emph{Renewable and Sustainable Energy Reviews}, vol. 157, p.
  112033, 2022.

\bibitem[Larsen et~al.(2023)Larsen, Christiansen, Ruepp, and
  Berger]{larsen2023toward}
L.~M. Larsen, H.~L. Christiansen, S.~Ruepp, and M.~S. Berger, ``Toward greener
  5g and beyond radio access networks—a survey,'' \emph{IEEE Open Journal of
  the Communications Society}, vol.~4, pp. 768--797, 2023.

\bibitem[Xu et~al.(2020)Xu, Zhou, Zhang, Wang, Liu, An, Shi, Liu, and
  Ma]{xu2020understanding}
D.~Xu, A.~Zhou, X.~Zhang, G.~Wang, X.~Liu, C.~An, Y.~Shi, L.~Liu, and H.~Ma,
  ``Understanding operational 5g: A first measurement study on its coverage,
  performance and energy consumption,'' in \emph{Proceedings of the Annual
  conference of the ACM Special Interest Group on Data Communication on the
  applications, technologies, architectures, and protocols for computer
  communication}, 2020, pp. 479--494.

\bibitem[Narayanan et~al.(2021)Narayanan, Zhang, Zhu, Hassan, Jin, Zhu, Zhang,
  Rybkin, Yang, Mao, et~al.]{narayanan2021variegated}
A.~Narayanan, X.~Zhang, R.~Zhu, A.~Hassan, S.~Jin, X.~Zhu, X.~Zhang, D.~Rybkin,
  Z.~Yang, Z.~M. Mao \emph{et~al.}, ``A variegated look at 5g in the wild:
  performance, power, and qoe implications,'' in \emph{Proceedings of the 2021
  ACM SIGCOMM 2021 Conference}, 2021, pp. 610--625.

\bibitem[Tombaz et~al.(2015)Tombaz, Frenger, Athley, Semaan, Tidestav, and
  Furuskar]{tombaz2015energy}
S.~Tombaz, P.~Frenger, F.~Athley, E.~Semaan, C.~Tidestav, and A.~Furuskar,
  ``Energy performance of 5g-nx wireless access utilizing massive beamforming
  and an ultra-lean system design,'' in \emph{2015 IEEE Global Communications
  Conference (GLOBECOM)}.\hskip 1em plus 0.5em minus 0.4em\relax IEEE, 2015,
  pp. 1--7.

\bibitem[Yu et~al.(2015)Yu, Chen, Yin, Zhang, and Li]{yu2015joint}
G.~Yu, Q.~Chen, R.~Yin, H.~Zhang, and G.~Y. Li, ``Joint downlink and uplink
  resource allocation for energy-efficient carrier aggregation,'' \emph{IEEE
  Transactions on Wireless Communications}, vol.~14, no.~6, pp. 3207--3218,
  2015.

\bibitem[Fu et~al.(2020)Fu, Soltani, Alshaer, Wang, Safari, McLaughlin, and
  Haas]{fu2020end}
Y.~Fu, M.~D. Soltani, H.~Alshaer, C.-X. Wang, M.~Safari, S.~McLaughlin, and
  H.~Haas, ``End-to-end energy efficiency evaluation for b5g ultra dense
  networks,'' in \emph{2020 IEEE 91st Vehicular Technology Conference
  (VTC2020-Spring)}.\hskip 1em plus 0.5em minus 0.4em\relax IEEE, 2020, pp.
  1--6.

\bibitem[Fog(1996)]{fog1996optimization}
A.~Fog, ``Optimization manual 4 instruction tables,'' \emph{Copenhagen
  University College of Engineering, Software Optimization Resources,
  http://www. agner. org/optimize (Jan. 5, 2010),(1996-2009)}, pp. 1--154,
  1996.

\bibitem[Hao et~al.(2018)Hao, Chen, Hu, Hossain, and Ghoneim]{hao2018energy}
Y.~Hao, M.~Chen, L.~Hu, M.~S. Hossain, and A.~Ghoneim, ``Energy efficient task
  caching and offloading for mobile edge computing,'' \emph{Ieee access},
  vol.~6, pp. 11\,365--11\,373, 2018.

\bibitem[Barhumi et~al.(2003)Barhumi, Leus, and Moonen]{barhumi2003optimal}
I.~Barhumi, G.~Leus, and M.~Moonen, ``Optimal training design for mimo ofdm
  systems in mobile wireless channels,'' \emph{IEEE Transactions on signal
  processing}, vol.~51, no.~6, pp. 1615--1624, 2003.

\end{thebibliography}
